\documentclass[times,twocolumn,final,authoryear]{elsarticle}
\usepackage[switch]{lineno}

\usepackage{amsmath}
\usepackage[hidelinks]{hyperref}

\journal{Advances in Space Research}

\usepackage{booktabs}

\biboptions{authoryear}

\begin{document}

\begin{frontmatter}

\title{A library of meteoroid environments encountered by \\ spacecraft in the inner solar system}

\author{Althea V.\ Moorhead}
\address{NASA Meteoroid Environment Office, Marshall Space Flight Center, Huntsville, AL 35812, USA}

\author{Katie Milbrandt}
\address{Department of Aerospace Engineering, Auburn University, Auburn, Alabama 36849, USA}

\author{Aaron Kingery}
\address{ERC, Inc., Marshall Space Flight Center, Huntsville, Alabama 35812}

\begin{abstract}
NASA's Meteoroid Engineering Model (MEM) is designed to provide aerospace engineers with an accurate description of potentially hazardous meteoroids. It accepts a spacecraft trajectory as input and its output files describe the flux, speed, directionality, and density of microgram- to gram-sized meteoroids relative to the provided trajectory. MEM provides this information at a fairly fine level of detail in order to support detailed risk calculations. However, engineers and scientists in the very early planning stages of a mission may not yet have developed a trajectory or acquired the tools to analyze environment data. Therefore, we have developed an online library\footnotemark\ of sample MEM runs that allow new users or overloaded mission planners to get a quick feel for the characteristics of the meteoroid environment. This library provides both visualizations of these runs and input files that allow users to replicate them exactly. We also discuss the number of state vectors needed to obtain an accurate representation of the environment encountered along our sample trajectories, and outline a process for verifying that any given trajectory is adequately sampled.
\end{abstract}


\end{frontmatter}

\section{Introduction}

In order to design reliable spacecraft, aerospace engineers must assess and mitigate a wide variety of risks.\footnotetext{https://fireballs.ndc.nasa.gov/mem/library/} A full understanding of all risks -- which can include phenomena such as spacecraft charging, corrosion, and impacts from orbital debris and meteoroids -- would require an impractically extensive knowledge of a wide variety of scientific and engineering disciplines. As a result, those conducting risk assessments must content themselves with a surface-level understanding of most fields and consult with experts when necessary. Primers and libraries, if well designed, can assist the risk assessment process by quickly familiarizing users with the basics of a field and reducing their dependence on external subject matter experts.

The members of NASA's Meteoroid Environment Office (MEO) often serve as subject matter experts on the topic of meteoroid impacts. We advise programs on the basic characteristics of the meteoroid environment; issue meteor shower forecasts and advisories \citep{2019JSpRo..56.1531M}; and develop and distribute the Meteoroid Engineering Model (MEM), a stand-alone piece of software that generates meteoroid environment data that is specific to a user's spacecraft trajectory \citep{2020JSpRo..57..160M,MEM3TM}. We have found that users or would-be users of MEM have widely varying levels of familiarity with the risk posed by meteoroid impacts and ballistic limit calculations. The missions which these users are planning can also vary dramatically in type and maturity. In many cases, users may not have a quantitative description of their spacecraft's trajectory, which makes early risk assessment very difficult.

In this paper, we present a library of sample spacecraft trajectories and corresponding MEM run results that is designed to help users conducting early risk assessments or learning to use the software. This library allows such users to quickly assess the typical meteoroid flux for various orbits and visualize the distribution of meteoroid speeds, directionality, and bulk particle density. In some cases, the user may find that one of our sample trajectories is similar to their own planned mission trajectory, and could potentially use our data files to conduct a preliminary risk assessment.

We do not construct original trajectories; instead, we sample known trajectories of existing spacecraft at a number of different locations in the inner solar system (see Section~\ref{sec:select}). We discuss our trajectory sampling in Section~\ref{sec:sample} and present a method for determining the trajectory resolution needed to attain a given precision in the meteoroid fluxes reported by MEM (Section~\ref{sec:res}).
Finally, we describe the visualizations and data provided in our online library and discuss their possible applications (Section~\ref{sec:data} and \ref{sec:summary}).

\section{Trajectory selection}
\label{sec:select}

We downloaded or generated trajectories for at least one spacecraft orbiting each planet in the inner solar system as well as the Moon. We have also included one trajectory that does not orbit any planet, but instead corresponds to the route by which the Mars Atmosphere and Volatile EvolutioN (MAVEN) spacecraft traveled from Earth to Mars.

Most of our selected trajectories are those of real US government (NASA or NOAA) spacecraft whose ephemerides are available through the Jet Propulsion Laboratory's Horizons ephemeris service.\footnote{https://ssd.jpl.nasa.gov/?horizons} We chose these particular spacecraft because they follow orbits that could be considered representative of a class of orbits: for instance, Aqua is an example of a satellite on a Sun-synchronous orbit. In one case, no suitable trajectory was available from Horizons: we therefore instead sampled a near-rectilinear halo orbit (NRHO) generated by \cite{lee19}.

\subsection{Earth-orbiting trajectories}

Most spacecraft orbit Earth; therefore, we have opted to generate a larger number of sample trajectories for Earth-orbiting spacecraft than for those orbiting other bodies.

\subsubsection{ISS: low Earth orbit}

We selected the International Space Station (ISS) as an example spacecraft in low Earth orbit (LEO). LEO describes orbits with altitudes within 2000~km of Earth's surface as well as the region of space where these orbits lie. This is a heavily populated region of space and is often chosen for communications or Earth observation satellites. 

\begin{figure}
    \centering
    \includegraphics{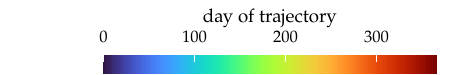}
    \includegraphics{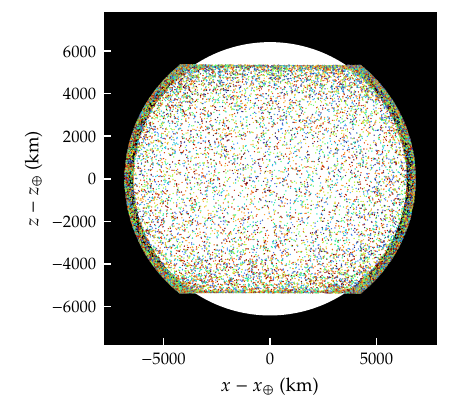}
    \caption{The ISS trajectory used for our library. The coordinate system is inertial and aligned with Earth's equator.}
    \label{fig:iss}
\end{figure}

The ISS orbits at an altitude just over 400~km above Earth's surface and has an orbital period between 90 and 93 minutes; its trajectory is depicted in Fig.~\ref{fig:iss}.\footnote{Additional views are available at\\ https://fireballs.ndc.nasa.gov/mem/library/iss.html}  Its altitude is small compared to Earth's radius, and this tends to protect the nadir (or Earth-facing) side of the spacecraft from meteoroid impacts (readers can jump ahead to Section~\ref{sec:bar} for a visualization). This effect is known as planetary shielding and is significant when the spacecraft is within a planetary radius of the planet's surface \citep{kessler72,jones07,moorhead20}.

\subsubsection{Aqua: Sun-synchronous orbit}

\begin{figure}
    \centering
    \includegraphics{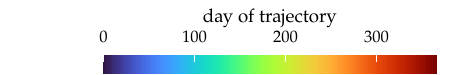}
    \includegraphics{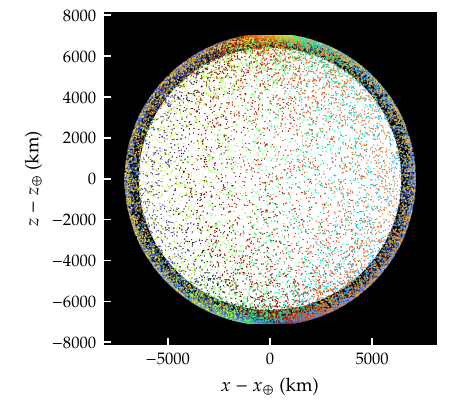}
    \caption{The Aqua trajectory used for our library. The coordinate system is inertial and aligned with Earth's equator.}
    \label{fig:aqua}
\end{figure}

We selected Aqua as an example spacecraft in Sun-synchronous orbit (SSO). SSO is a type of near-polar LEO with an altitude and inclination chosen such that the orbit precesses around Earth once per year. This tends to preserve the angle between the orbit and the Sun-Earth vector. Such satellites will therefore always view locations on Earth's surface at the same illumination angle. This orbit is popular with imaging and weather satellites. Aqua's trajectory is depicted in Fig.~\ref{fig:aqua}.

Aqua is part of NASA's Earth Observing System (EOS). It has an orbital altitude of about 702-703~km and an orbital period of 99 minutes. Like the ISS, Aqua's nadir surface will be largely shielded from the meteoroid environment by Earth.

\subsubsection{GOES-14: Geostationary orbit}

A geostationary or geosynchronous equatorial orbit (GEO) is one in which the spacecraft's mean motion matches Earth's rotational period. A spacecraft in GEO orbits Earth exactly once per sidereal day and thus is always positioned directly above the same point on Earth's surface. GEO orbits are used by one-way communications and weather satellites, but cannot communicate with latitudes more than 81$^\circ$ from the equator.

\begin{figure}
    \centering
    \includegraphics{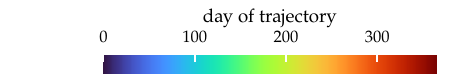}
    \includegraphics{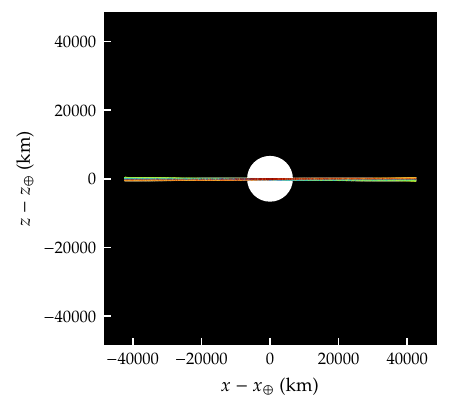}
    \caption{The GOES-14 trajectory used for our library. The coordinate system is inertial and aligned with Earth's equator.}
    \label{fig:goes14}
\end{figure}

The GOES-14 satellite is part of the National Oceanic and Atmospheric Administration (NOAA)'s Geostationary Operational Environmental Satellite (GOES) system. It generally maintains a geographical longitude of ${\sim 105^\circ}$W. With a standard GEO altitude of $35,786$~km (5.6 Earth radii), GOES-14 benefits very little from planetary shielding. The total flux on the spacecraft varies very little, although the apparent directionality changes as the spacecraft changes its orientation relative to Earth's direction of motion about the Sun. The trajectory is depicted in Fig.~\ref{fig:goes14}.

\subsubsection{JWST: Sun-Earth L2}

\begin{figure}
    \centering
    \includegraphics{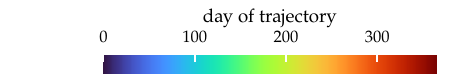}
    \includegraphics{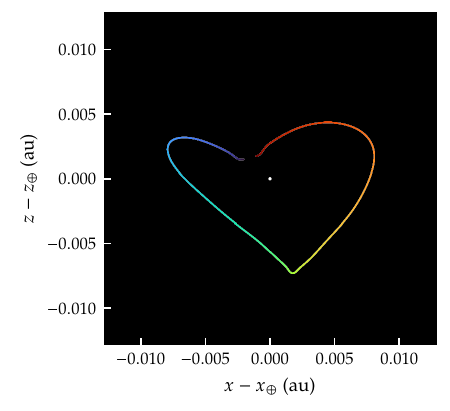}
    \caption{The JWST trajectory used for our library. The coordinate system is inertial and aligned with Earth's equator.}
    \label{fig:jwst}
\end{figure}

The James Webb Space Telescope (JWST) is a large, multipurpose infrared space observatory. It was launched in late 2021 and its first images were released in July 2022. The spacecraft follows a halo orbit around the Sun-Earth L2 Lagrange point and thus remains near Earth. The trajectory is depicted in Fig.~\ref{fig:jwst}. 

It would be incorrect to say that JWST follows an interplanetary \emph{trajectory}, as it does not travel between planets. However, JWST does not orbit Earth and will not ``see'' the effects of Earth on the meteoroid environment. We have therefore placed JWST in this section because it lies at the edge of Earth's sphere of gravitational influence and the environment it encounters is best described as interplanetary (that is, it is relatively unaffected by gravitational focusing and planetary shielding). JWST's orbital motion about L2 is quite slow and therefore we expect it to encounter a fairly constant flux.

\subsection{Moon-orbiting trajectories}

\subsubsection{LADEE: low lunar orbit}

\begin{figure}
    \centering
    \includegraphics{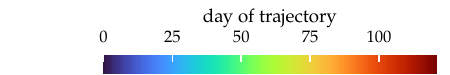}
    \includegraphics{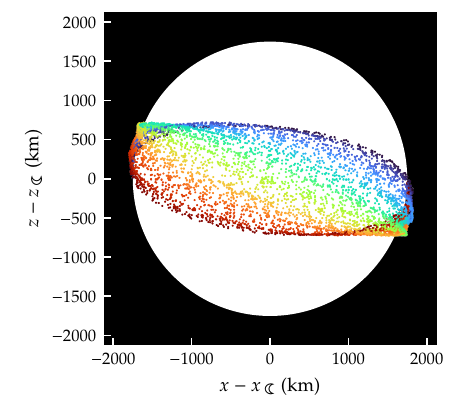}
    \caption{The LADEE trajectory used for our library. The coordinate system is inertial and aligned with the Moon's equator.}
    \label{fig:ladee}
\end{figure}

The Lunar Atmosphere and Dust Environment Explorer (LADEE) was a NASA mission to study the lunar exosphere. 
It orbited the Moon for just over 6 months in a retrograde low lunar orbit.
In general, lunar orbits are unstable and it is apparent from LADEE's ephemeris that the spacecraft adjusted its orbital period approximately 20 times in a 6 month period. The trajectory is depicted in Fig.~\ref{fig:ladee}.

LADEE's altitude ranged from just a few km above the lunar surface to 148~km. We therefore expect its nadir surfaces to be fairly protected from the meteoroid environment; note, however, that dust particles originating from the lunar surface are not modeled in MEM. Users will need to use a separate model for these additional particles \citep[see][and references therein]{memosee,sp8013}.

\subsubsection{Near-rectilinear halo orbit}

A near-rectilinear halo orbit (NRHO) is being considered for NASA's Lunar Gateway. This orbit is so-named because its orbital eccentricity relative to the Moon varies, and some portions of the trajectory are therefore straighter than that of an ellipse.

\begin{figure}
    \centering
    \includegraphics{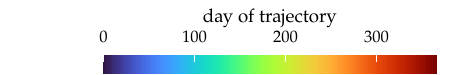}
    \includegraphics{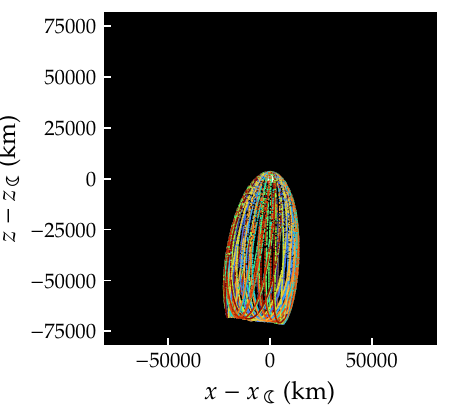}
    \caption{The NRHO trajectory used for our library. The coordinate system is inertial and aligned with the Moon's equator.}
    \label{fig:nrho}
\end{figure}

No spacecraft have used a lunar NRHO trajectory in the past, and we therefore cannot use our usual approach of downloading a sample trajectory from Horizons. Instead, we extracted state vectors using the spy utility\footnote{https://naif.jpl.nasa.gov/naif/utilities.html} from a binary Spice file prepared by \cite{lee19}. The trajectory is depicted in Fig.~\ref{fig:nrho}.

In our example NRHO, the spacecraft circles the Moon, which in turn orbits Earth, which in turn orbits the Sun. In order to fully cover all possible Sun-Earth-Moon-spacecraft positions, we would need to sample over a 19-year Metonic cycle. However, an object on an NRHO spends most of its time at large selenocentric distances, moving slowly relative to the Moon. The overall meteoroid environment is therefore fairly stable and we thus restrict our analysis to a one-year period.

Please note that there is a ``family'' of NRHO orbits around any L1, L2, or L3 Lagrange point. This orbit and the corresponding meteoroid environment should be considered illustrative of the Earth-Moon L2 NRHO family; the environment encountered along another member of the orbit family will differ from that in our library.

\subsection{Orbits around other planets}

MEM~3 accepts trajectories with heliocentric distances between 0.2 and 2~au; within this range, it automatically detects nearby planets and accounts for their gravitational influence on the meteoroid environment. Thus, the software is capable of handling trajectories of spacecraft orbiting Mercury, Venus, and Mars.

\subsubsection{MESSENGER: Mercurian orbit}

\begin{figure}
    \centering
    \includegraphics{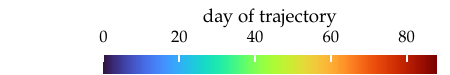}
    \includegraphics{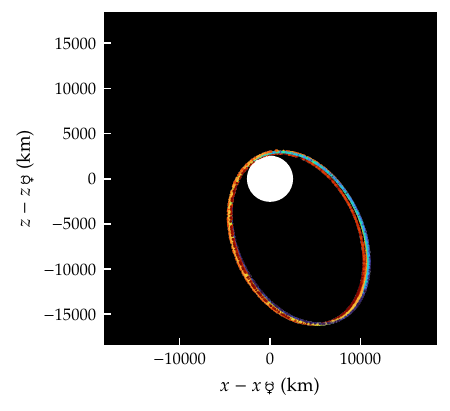}
    \caption{The MESSENGER trajectory used for our library. The coordinate system is inertial and aligned with Mercury's equator.}
    \label{fig:messenger}
\end{figure}

MESSENGER (Mercury Surface, Space Environment, Geochemistry, and Ranging) was a NASA mission to study Mercury's surface and magnetic field. It orbited Mercury for four Earth years, maintaining an elliptical orbit that limited the spacecraft's exposure to the hot Hermean surface. Its trajectory is depicted in Fig.~\ref{fig:messenger}.

Mercury orbits the Sun in 88 days; we therefore chose to sample the first 88 days of MESSENGER's primary mission. The spacecraft's orbit does not precess in an inertial frame; thus, one Mercury orbit covers the full range of Sun-Mercury-MESSENGER relative positions.

\subsubsection{Venus Express: Venusian orbit}

\begin{figure}
    \centering
    \includegraphics{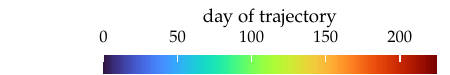}
    \includegraphics{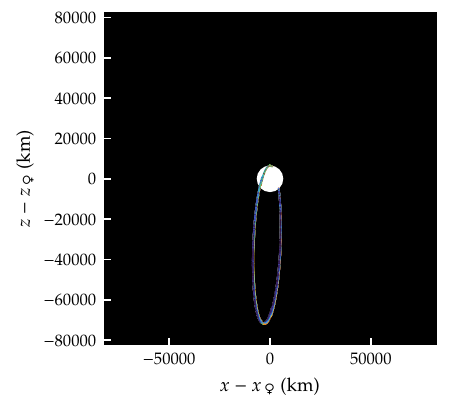}
    \caption{The Venus Express trajectory used for our library. The coordinate system is inertial and aligned with Venus's equator.}
    \label{fig:venusexp}
\end{figure}

Venus Express was a European Space Agency (ESA) mission to study the atmosphere of Venus. It used a polar, elliptical orbit for its primary mission and most of its extended missions. Its trajectory is depicted in Fig.~\ref{fig:venusexp}

Venus orbits the Sun in 225 days; we therefore sampled the first 225 days of Venus Express's mission. Like MESSENGER, Venus Express's orbit does not precess relative to the planet in an inertial frame.

\subsubsection{MRO: low Martian orbit}

\begin{figure}
    \centering
    \includegraphics{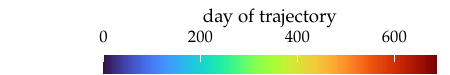}
    \includegraphics{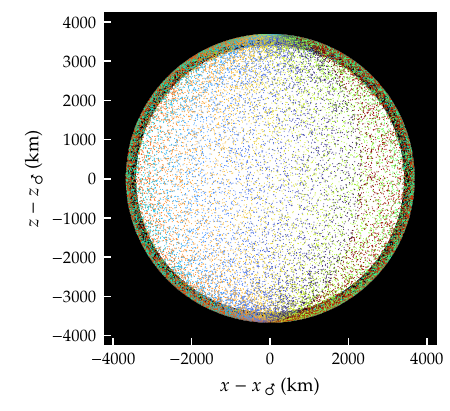}
    \caption{The MRO trajectory used for our library. The coordinate system is inertial and aligned with Mars's equator.}
    \label{fig:mro}
\end{figure}

NASA's Mars Reconnaissance Orbiter (MRO) has been studying the Martian surface since 2006 and is still operational. Unlike MESSENGER and Venus Express, MRO remains within a few hundred km of the Martian surface on a nearly circular orbit. Its near-polar orbit precesses once per Martian year; MRO is therefore in a Martian Sun-synchronous orbit. We have sampled the first 687 days (one Martian year) of MRO's primary science phase. The trajectory is depicted in Fig.~\ref{fig:mro}.

\subsubsection{MAVEN: Martian orbit}

NASA's Mars Atmosphere and Volatile Evolution (MAVEN) orbiter arrived at Mars in late 2014. After surviving an encounter with comet C/2013 A1 (Siding Spring), MAVEN begin primary science operations in November of that year.

\begin{figure}
    \centering
    \includegraphics{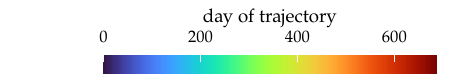}
    \includegraphics{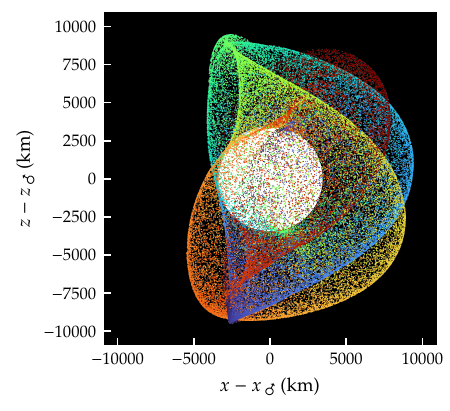}
    \caption{The MAVEN trajectory used for our library. The coordinate system is inertial and aligned with Mars's equator.}
    \label{fig:maven}
\end{figure}

MAVEN maintains an eccentric orbit that is highly inclined but not polar. Within its first Martian year (687 Earth days), we see an irregular drift in the orbital period and a fairly complex distribution of spacecraft locations (see Fig.~\ref{fig:maven}).

\subsection{Interplanetary space}

\subsubsection{MAVEN: Earth-to-Mars transfer}

As mentioned above, MEM automatically determines whether any massive bodies lie near the user's trajectory. While MEM permits the user to select any of the inner solar system planets as the origin of the input trajectory, it does not require that the user place the origin at the nearest planet. This flexibility in selecting the coordinate origin is particularly useful for trajectories in which a spacecraft travels from one planet (or the Moon) to another.

\begin{figure}
    \centering
    \includegraphics{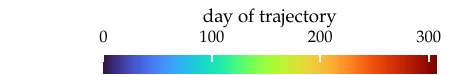}
    \includegraphics{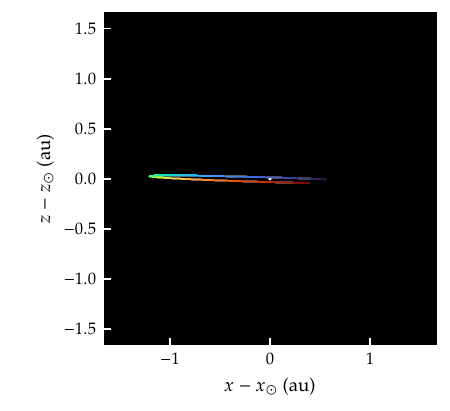}
    \caption{The MAVEN transfer trajectory used for our library. The coordinate system is inertial and aligned with the ecliptic.}
    \label{fig:transfer}
\end{figure}

We use MAVEN once again for our example trajectory. In this case, we follow MAVEN from its launch from Earth to its arrival at Mars (see Fig.~\ref{fig:transfer}). We use Barycentric Dynamical Time (TDB) in all our trajectory files, but this is particularly critical for a transfer trajectory. The TDB time scale increases in a predictable fashion, unlike Coordinated Universal Time (UTC), which is frequently updated to include leap seconds. Currently, the two timescales differ by 69 seconds. When the nearest massive body differs from the origin of the input trajectory's coordinate system, however, it is important that the user provide times in TDB so that MEM can correctly compute the distance to the nearest massive body.

For most of this portion of its trajectory, MAVEN is fully exposed to the interplanetary meteoroid environment. The meteoroid flux decreases with heliocentric distance, and therefore the rate of impacts onto MAVEN decreased as it moved away from the Sun.

\section{Trajectory sampling}
\label{sec:sample}

We used the Horizons application programming interface (API) to sample state vectors for our selected spacecraft. In general, we opted to sample over either [1] the length of time that the orbited body requires to orbit the Sun once (ISS, Aqua, GOES-14, NRHO, MESSENGER, Venus Express, MRO, MAVEN), [2] the length of time that the spacecraft itself requires to orbit the Sun (JWST), or [3] the duration of the mission (LADEE, MAVEN transfer). The exact intervals are provided in Table~\ref{tab:summ}.

\begin{table*}
\centering
\begin{tabular}{l@{\hspace{24pt}}ll@{\hspace{24pt}}lr@{\,--\,}lr@{~}lr} \hline \hline \addlinespace[2pt]
&
    interval & interval &
    nearest \\
&
    start & end &
    body & 
    \multicolumn{2}{c}{altitude} &
    \multicolumn{2}{c}{period} &
    inc. \\ \addlinespace[2pt] \hline \addlinespace[2pt]
ISS & 
    2020 Jan 1 & 2021 Jan 1 &
    Earth & 415 & 435 km & 
    93 & min & 52$^\circ$ \\
Aqua & 
    2020 Jan 1 & 2021 Jan 1 &
    Earth & 700 & 719 km & 
    99 & min & 98$^\circ$ \\
GOES-14 & 
    2015 Jan 1 & 2016 Jan 1 &
    Earth & 35,765 & 35,821 km & 
    23.93 & hr & 0$^\circ$ \\
JWST & 
    2023 Jan 1 & 2024 Jan 1 &
    Earth & 0.0082 & 0.0117 au & 
    6 & mos. & *\hphantom{$^\circ$} \\
LADEE & 
    2013 Dec 8 & 2014 Apr 8 &
    Moon & 5 & 148 km & 
    115 & min & 157$^\circ$ \\
NRHO & 
    2021 Jan 01 & 2022 Jan 02 &
    Moon & 1,481 & 70,076 km & 
    8.67 & d & *\hphantom{$^\circ$} \\
MESSENGER & 
    2011 Apr 4 & 2011 Jul 1 &
    Mercury & 226 & 15,166 km & 
    12.03 & hr & 83$^\circ$ \\
Venus Express & 
    2006 May 8 & 2006 Dec 19 &
    Venus & 269 & 66,667 km & 
    24.03 & hr & 90$^\circ$ \\
MRO & 
    2006 Nov 7 & 2008 Sep 24 &
    Mars & 233 & 307 km & 
    112 & min & 93$^\circ$ \\
MAVEN & 
    2014 Nov 16 & 2016 Oct 3 &
    Mars & 113 & 6497 km & 
    4.5 & hr & 74$^\circ$ \\[4pt]
MAVEN & 
    2013 Nov 18 & 2014 Sep 22 &
    Sun & 0.97 & 1.45 au & 
    \multicolumn{2}{c}{\emph{na}} & $5^\circ$ \\
(transfer) &
    19:22 TDB & 02:24 TDB \\
\end{tabular}
\caption{Spacecraft trajectories selected for our run library. We provide the start and end times; the nearest massive body; and the orbital altitude, period, and inclination relative to that body. Inclination is measured relative to the nearest body's equator, if that body is a moon or planet, or the ecliptic, if that body is the Sun.\\[6pt]
$^*$JWST's orbit and the NRHO are both nearly perpendicular to the closest body's \emph{orbital plane}, not its equator.}
\label{tab:summ}
\end{table*}

\begin{figure}[tp]
    \centering
    \includegraphics{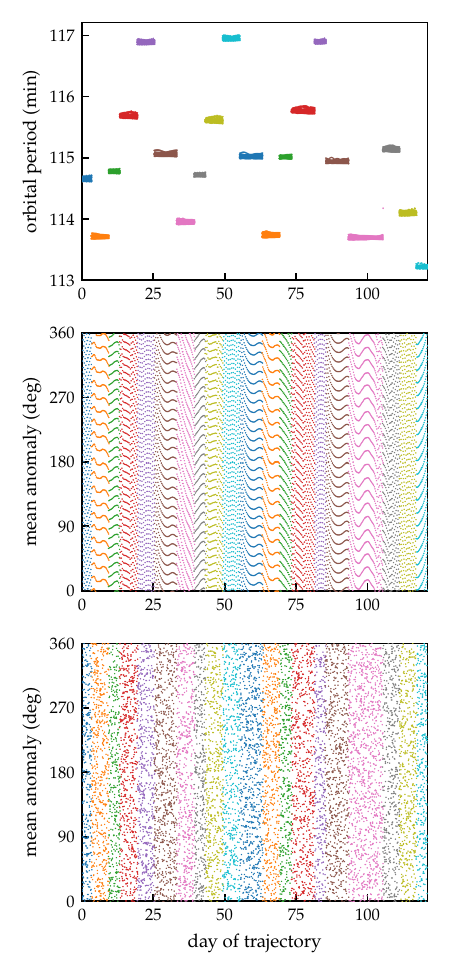}
    \caption{Orbital period (top) and mean anomaly of the LADEE spacecraft relative to the Moon, sampled at regular 17-minute intervals (center) and at random times (bottom).}
    \label{fig:ladcad}
\end{figure}

Initially, we attempted to use a constant sampling cadence, taking care to select a cadence that does not evenly divide the orbital period. For instance, the ISS has an orbital period of about 93 minutes and we selected a sampling cadence of 17 minutes. This would result in only a few samples from each of the 5000+ orbits ISS completes in a year, but the mean anomalies sampled would vary from one orbit to the next.
However, this approach proved to be troublesome in many cases: for instance, LADEE changed its orbital period frequently during the course of its mission. We found that any constant sampling cadence we tried resulted in periodicities in the sampled mean anomaly (see Fig.~\ref{fig:ladcad}).

We decided to sidestep the orbit sampling interval issue by instead sampling at random times within the chosen interval. We set the number of samples to correspond to the length of the time interval divided by our original cadence choice of 17 minutes. We rounded the dates to the nearest $10^{-6}$ day (less than 0.1~s); if two dates rounded to the same value, we discarded one and drew a new random date. We then used the Horizons API to download the trajectory data corresponding to these times in batches of 50.

We downloaded data in a coordinate system that is aligned with the International Celestial Reference Frame (ICRF). Thus, the $x$-$y$ plane is parallel to Earth's mean equator at the J2000 epoch. We set the coordinate center to be that of the orbited body (Earth for ISS, Mars for MRO, etc.). We made one exception: the MAVEN transfer trajectory was downloaded in ecliptic coordinates, with the Sun at the origin. For spacecraft orbiting bodies other than Earth, we also downloaded the same trajectory relative to what Horizons calls the ``body mean equator and node of date'' reference plane. We use this latter coordinate frame for orbit visualizations, but not as MEM input.

Prior to using any trajectory, we verify that the orbit is evenly sampled using a chi-squared test on either mean anomaly (for orbits with $\text{max}(e) > 0.02$) or the angle from the equator (for orbits with $\text{max}(e) < 0.02$). Here, $e$ refers to the orbital eccentricity, and the angle from the equator can be computed as:
\begin{align}
    \omega + f &= \sin^{-1} \left( \frac{z}{r \sin i} \right)
\end{align}
where $\omega$ is the argument of pericenter, $f$ is true anomaly, $z$ is the vertical offset from the equator of the orbited body, $r$ is the distance from the center of the orbited body, and $i$ is the orbital inclination.

We rounded our randomly generated Julian dates to 6 significant figures; we took care to do this rounding prior to submitting them to the Horizons API. We output dates in Barycentric Dynamical Time (TDB) and the corresponding state vectors in units of km and km~s$^{-1}$, as required by MEM.

\section{Trajectory resolution}
\label{sec:res}

In order to use MEM, users must convert their trajectory to a series of Julian dates and state vectors. If an orbit or trajectory is sampled too infrequently, the resulting environment description may be inaccurate or noisy. On the other hand, if the trajectory is sampled too finely, the user may find themselves waiting days or even weeks for their MEM run to finish (see Table~\ref{tab:runtimes} for approximate run times). Unfortunately, we cannot provide a single recommended minimum number of state vectors; the needed number will depend on the trajectory and the user's needs.

\begin{table}
    \centering
    \begin{tabular}{ll} \hline \hline \addlinespace[2pt]
        trajectory size &  run time \\ \addlinespace[2pt] \hline \addlinespace[2pt]
        100 & 2-3 min \\
        1000 & 20-30 min \\
        10,000 & 4 hr
    \end{tabular}
    \caption{Approximate run times (right column) for different input trajectory sizes, given in terms of number of state vectors (left column). These estimates were obtained by running an ISS-like trajectory in high-fidelity mode on a laptop with a 2.4~GHz processor. Run times can differ depending on the user's selected run options and computer specifications.}
    \label{tab:runtimes}
\end{table}

To help mitigate this, MEM offers a ``random draw'' option; when selected, MEM analyzes only a random sample of state vectors from the user's input file. This option can be used to complete short test runs or to investigate whether results might converge for a smaller number of state vectors.

MEM can also output files that provide the standard deviation of the flux along the trajectory. This option can be used in conjunction with a short random draw run to predict the number of state vectors needed for convergence.

\begin{figure}
    \centering
    \includegraphics{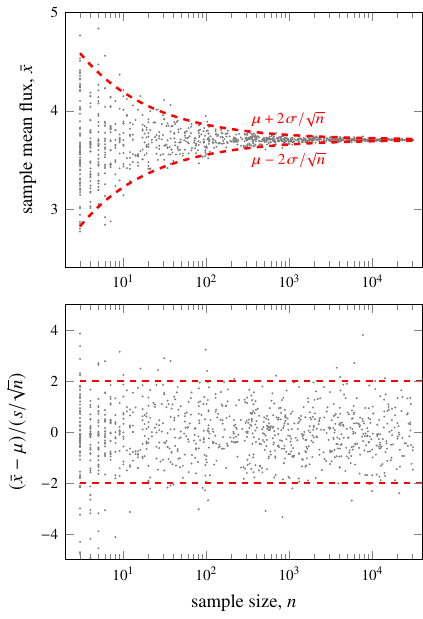}
    \caption{The mean flux of high-density particles relative to our ISS trajectory. We generated this plot by saving the total flux for each state vector in the trajectory, then drawing random samples of varying size from the data and calculating the sample mean and standard deviation. We find these quantities behave as expected; for sample sizes larger than 30, approximately 95\% of sample means fall within two sigma of the population mean.}
    \label{fig:clt}
\end{figure}

The central limit theorem for sample means \citep[see, e.g., Sec.~5.2 of][]{lock5} predicts that the mean of a sample of size $n$ drawn from a population with mean $\mu$ and standard deviation $\sigma$ will tend to follow a normal distribution with mean $\mu$ and standard deviation:
\begin{align}
    \sigma_m &= \frac{\sigma}{\sqrt{n}}
    \label{eq:clt}
\end{align}
Furthermore, the sample mean ($\bar{x}$) and variance ($s^2$) are unbiased estimators for the population mean and variance:\footnote{This does not hold for the standard deviation, $s$; the unbiased estimator for the population standard deviation includes an additional correction factor of $\sqrt{2/(n-1)} \, \Gamma \mathopen{}\left( \frac{n}{2} \right)\mathclose{} \, / \, \Gamma \mathopen{}\left( \frac{n-1}{2} \right)\mathclose{}$ \citep{bolch68}. However, for sample sizes greater than 100, this correction factor adjusts the estimate by less than 0.25\% and we therefore omit it from our equations.}
\begin{align}
    \bar{x} &= \frac{1}{n} \sum_{i=1}^n x_i \\
    s^2 &= \frac{1}{n-1} \sum_{i=1}^n (x_i - \bar{x})^2
\end{align}
These equations can be combined and inverted to estimate the needed sample size, $n$. This behavior is illustrated in Fig.~\ref{fig:clt}.

\subsection{Method for estimating trajectory sample size}
\label{sec:method}

We suggest the following process for determining the needed sample size. 

First, complete a random-draw run by setting the desired random draw size to 100 state vectors; a run of this size should take minutes, not hours. Be sure to select the standard deviation file output option so that the sample standard deviations are saved to file.

\begin{figure}
    \centering
    \includegraphics[width=\linewidth]{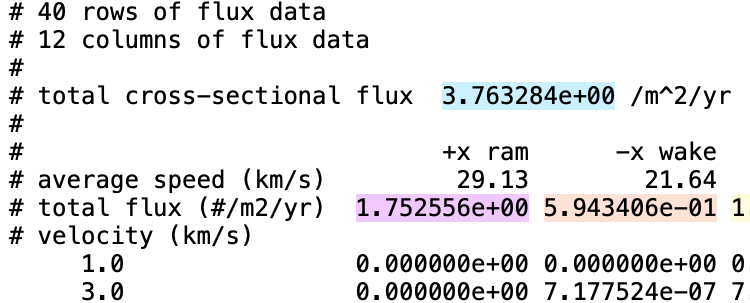}
    \caption{A portion of a MEM output file -- cube\_avg.txt -- in which certain fluxes have been highlighted. In the fourth row, we highlight the ``total flux'': this is the flux of meteoroids onto an object with no fixed orientation per unit mean cross-sectional area. In the eighth row, we highlight the average flux on the surfaces of a cube with a ``body-fixed'' orientation. The full file contains the flux on all six faces of the cube, plus the flux on surfaces facing Earth, Sun, and directly away from the Sun. The file cube\_std.txt has exactly the same format, but all quantities are standard deviations rather than means.}
    \label{fig:cubefile}
\end{figure}

Second, identify the flux quantities of interest \citep[see Sec.~3.4 of the MEM user guide for a description of all such quantities;][]{MEM3TM}. If the user is uncertain about which quantities to select, we suggest using the so-called ``cube fluxes'' provided in ``cube\_avg.txt'' and ``cube\_std.txt.'' (There are two such sets of files for each run; one for the high-density population of meteoroids and another for low-density meteoroids.) An example is shown in Fig.~\ref{fig:cubefile}. We will denote the value or values from the average flux file as $\bar{x}$ or $\bar{x}_j$, while the value or values from the standard deviation file will be denoted $s$ or $s_j$. Here, $j$ is an optional index used to distinguish between multiple flux components of interest.

Third, select the desired accuracy for the final run ($p$) and the acceptable probability that the final run does not satisfy that accuracy requirement ($\alpha$). Select or calculate the $z$-statistic that corresponds to your chosen $\alpha$ value using the 68-95-99.7 rule, a z-table (e.g., see Table~\ref{tab:alpha}) or the following equation:
\begin{align}
    z &= \sqrt{2} \, \text{erf}^{-1} (1 - \alpha)
\end{align}
where $\text{erf}^{-1}$ is the inverse error function.

\begin{table}
    \centering
    \begin{tabular}{lc} \hline \hline \addlinespace[2pt]
        $\alpha$ & $z$ \\ \addlinespace[2pt] \hline \addlinespace[2pt]
         31.7\% & 1 \\
         4.55\% & 2 \\
         0.27\% & 3 
    \end{tabular}
    \caption{Three commonly used confidence levels ($\alpha$) and their corresponding $z$-values.}
    \label{tab:alpha}
\end{table}

Finally, estimate the number of state vectors needed to obtain the desired accuracy ($p$) for the $j$\textsuperscript{th} quantity at the desired level of confidence ($1-\alpha$): 
\begin{align}
    n &\ge n_{\sigma, j} = \left( \frac{z \, s_j}{p \, \bar{x}_j} \right)^2
    \label{eq:nmin}
\end{align}
If multiple quantities are used -- such as the total flux from each population, or the flux on each surface listed in the ``cube'' file -- the sample size should be the largest value of $n_j$.

The user can also consider requiring convergence to within a constant value that is 100$p$\% of the \emph{largest} flux element under consideration. This is equivalent to replacing $\bar{x}_j$ with $\sup_j \bar{x}_j$ in the denominator of Eq.~\ref{eq:nmin}. This helps the user avoid situations in which an element with a small flux but large relative variation drives the sample size to large numbers. For instance, when a spacecraft in low Earth orbit keeps the same surface pointed towards Earth, the flux of meteoroids on that surface will be quite low, and it may not be desirable to have results converge to within $p$ of a negligible flux.

If the calculated sample size is greater than 100, the user can use either re-sample their trajectory to create a file with the required number of state vectors, or they can use MEM's built-in random draw option to specify the sample size (if the size of the existing trajectory file exceeds the sample size). Please note that MEM runs in high-fidelity mode when generating standard deviation files. Thus, the user should also use the high-fidelity option for their final run. If the calculated sample size is less than 100, the user can proceed to use their 100-state-vector results; there is no need to generate a smaller, coarser run.

\subsection{Sample size and skewness}
\label{sec:skew}

The central limit theorem does \emph{not} require that the underlying distribution be normal. However, if the underlying distribution is non-normal, there is a sample size, usually given as 30, at which the sample is considered ``large enough'' for Eq.~\ref{eq:clt} to apply. The bottom panel of Fig.~\ref{fig:clt} hints at this requirement: when $n < 10$, there is a wider spread in the residuals.

However, 30 state vectors may not be ``large enough'' if the distribution is extremely skewed. Unfortunately, we were unable to locate any quantitative guideline as to what constitutes a sufficiently large sample size as a function of skewness. We therefore decided to construct our own, empirical formula.

We used the 104 univariate continuous distributions available in SciPy version 1.10.0 \citep{scipy20} as the basis for our experiment. For each distribution, we randomly generated 50 sets of positive shape parameters ${\lbrace q_0, ..., q_m \rbrace }$ using:
\begin{align}
    q_i &= - \ln u_i
\end{align}
where $u_i$ is a uniform random variate in the range ${[0, 1]}$. The parameter count $m$ is determined by the shape attribute of the distribution.

Not all distributions are compatible with our approach. We excluded those whose shape attribute is ``None'' -- that is, whose only shape parameters are a location and a scale -- and those that consistently produced errors when used with our random parameters (possibly due to a restriction in range of allowed parameters). We also excluded those that required longer than 0.1~s to generate 1000 random variates.

\begin{figure}
    \centering
    \includegraphics{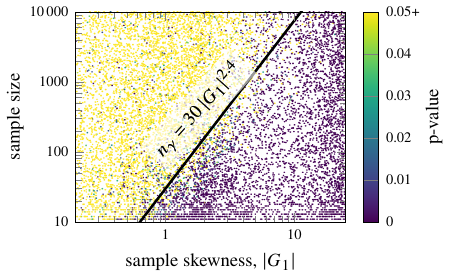}
    \caption{The p-value (color scale) for a test for normality of the means of 1000 samples of the given size and skewness. A p-value less than 0.05 is typically interpreted as evidence that the data deviate from a normal distribution. Thus, the points in light yellow correspond to sets of samples whose means are consistent with a normal distribution, and the darker, green-to-purple points correspond to sets of samples whose means are inconsistent with a normal distribution.}
    \label{fig:skew}
\end{figure}

For each distribution and set of random shape parameters, we first generated 1000 random variates and estimated the skewness using the  sample skewness (also known as the adjusted Fisher-Pearson skewness coefficient):
\begin{align}
    G_1 &= \frac{n}{(n-1)(n-2)} \sum_{i=1}^n \left(
        \frac{x_i - \bar{x}}{s}
    \right)^3 
\end{align}
We then generated a second set of random variates with the same distribution and shape parameters, but now with a random sample size, $n_\text{samp}$ between 5 and 10,000. We repeated this random drawing with the same sample size 1000 times and computed the mean of each sample. Finally, we performed a test for normality \citep{dagostino71,dagostino73} and recorded the p-value (see Fig.~\ref{fig:skew}).

We find that 
\begin{align}
    n &\ge n_\gamma = 30 \, |G_1|^{2.4}
    \label{eq:ngam}
\end{align}
appears to be a reasonable predictor of whether the sample means follow a normal distribution and thus the central limit theorem applies; the black line in Fig.~\ref{fig:skew} corresponds to this equation. We stress that this is an empirical rule and we cannot guarantee that it applies outside of the sampled range or to all possible univariate distributions. 

MEM does not currently report the sample skewness, although users can compute it themselves by choosing the ``intermediate files'' option. These intermediate files report all flux outputs for each state vector used.

\subsection{Application to example trajectories}

In this section, we apply the methods outlined in Sections~\ref{sec:method} and \ref{sec:skew} to the trajectories in our library. In each case, we performed an initial MEM run using a random sample of 100 state vectors from the master trajectory. We then use equations~\ref{eq:nmin} and \ref{eq:ngam} to estimate the number of state vectors needed for the central limit theorem to apply and for the reported flux values to lie within 1\% of that of the entire trajectory with 68\% confidence (i.e., we select $p=0.01$ and $z=1$).

We apply these criteria not only to the overall flux on the spacecraft, but to the surface fluxes reported in the ``cube\_avg.txt'' files for a ``body-fixed'' output frame. However, those spacecraft in low orbits around Earth or other bodies will have a much lower flux on their nadir-facing surface. We'd like to avoid scenarios in which the estimated sample size is inflated by requiring 1\% precision on a negligible flux; therefore, we instead require that all fluxes converge to 1\% of the \emph{maximum} surface flux:
\begin{align}
    n_{\sigma, j} &= \left( \frac{s_j}{\delta} \right)^2, ~ \text{where} ~ \delta = 0.01 \, \sup_j \bar{x}_j 
    \label{eq:nsig}
\end{align}
We also use Eq.~\ref{eq:ngam} to compute $n_{\gamma, j}$ for each surface.

Table~\ref{tab:issnums} reports $n_{\sigma,j}$ and $n_{\gamma,j}$ for the ISS total and surface fluxes due to both of MEM's meteoroid populations (i.e., the high- and low-density populations). In this case, we find that the number of state vectors required for the assumption of normality never exceeds the canonical recommendation of 30; this is not, however, true for every spacecraft. The largest sample size requirement in the table is 2883 and this is therefore the number of state vectors sampled for the ISS example in our run library.

\begin{table*}[]
    \centering
    \begin{tabular}{llrrrrrrrrrr} \hline \hline \addlinespace[2pt]
    && total & $+x$ & $-x$ & $+y$ & $-y$ & $+z$ & $-z$ & +Earth & +Sun & -Sun \\ \addlinespace[2pt] \hline \addlinespace[2pt]
    high density & $n_{\sigma, j}$ & 
        380 & 1680 &  637 &  832 &  783 &  655 &   21 &   21 & 2883 & 2411 \\
     & $n_{\gamma, j}$ & 
        1 &    1 &    9 &    5 &    5 &    1 &   25 &   25 &    2 &    1 \\[4pt]
    low density & $n_{\sigma, j}$ & 
        540 &  312 &  108 &  276 &  230 &  317 &    2 &    2 &  182 &  102 \\
    & $n_{\gamma, j}$ &  
        1 &    5 &   16 &    2 &    1 &    2 &    6 &    6 &    2 &    5 \\
    \end{tabular}
    \caption{The number of ISS state vectors required for convergence ($n_{\sigma,j}$) of the total flux, the flux on the surfaces of a cube in the VNB coordinate system (see Sec.~\ref{sec:bar}), and the flux on surfaces towards or away from Earth and Sun. The desired precision in the total flux is 1\% of its value, while the desired precision in the surface fluxes is 1\% of the maximum surface flux. In all cases, the acceptable significance level has been set to $\alpha = 31.7$\% (or $z = 1$). We also provide the number of state vectors required for assumption of normality ($n_{\gamma, j}$). The two density populations in MEM are listed separately.}
    \label{tab:issnums}
\end{table*}

In Table~5, we provide the recommended number of state vectors for each spacecraft as well as the surface and meteoroid population that drives that recommendation. For instance, we list $n=2883$, ``Sun,'' and ``high'' for ISS because the largest number in Table~\ref{tab:issnums} is 2883, and that number is derived from the flux of high-density meteoroids on a Sun-facing surface.

\begin{table}
    \centering
    \begin{tabular}{lrll} \hline \hline \addlinespace[2pt]
    spacecraft & $n$ & surface & population \\ \addlinespace[2pt] \hline \addlinespace[2pt]
    ISS       & 2883 & Sun      & high den.\\ 
    Aqua      & 2929 & anti-Sun & high den.\\
    GOES-14   &  947 & $-z$     & high den.\\
    JWST      & 1414 & $-z$     & high den.\\
    LADEE     & 4932 & anti-Sun & high den.\\
    NRHO      &  844 & $+y$     & high den.\\
    Messenger & 1351 & $-z$     & high den.\\
    VenusExp  & 1160 & anti-Sun & high den.\\
    MRO       & 3180 & anti-Sun & high den.\\
    MAVEN     & 1908 & $+x$     & high den.\\
    transfer  & 1488 & total    & low den.\\
    \end{tabular}
    \caption{The number of state vectors required for the convergence of the total flux to within 1\% of its value and the convergence of all surface fluxes to within 1\% of the largest surface flux. The third column specifies which flux component determined the value of $n$. In all cases, the largest value of $n$ arose from the high-density meteoroid population in MEM. The significance level has been set to $\alpha = 31.7$\% (or $z = 1$).}
    \label{tab:allnums}
\end{table}

In almost all cases, the recommended number of state vectors is driven by the precision requirement (Eq.~\ref{eq:nsig}). However, we did encounter an exception: our 100-state vector run for Venus Express had a high skewness value for Sun-facing and anti-Sun-facing surfaces (${\lvert G_1 \lvert = 4.6}$ and 4.2, respectively). A skewness of 4.6 requires over 1000 state vectors for the central limit theorem to apply. This appears to arise because Venus Express's orbit is eccentric and does not precess in a Sun-synchronous fashion. Therefore, the orientation of a Sun-facing surface changes with respect to both the spacecraft's orbit about Venus. At times, it is heavily protected by planetary shielding, and at others, it is both completely unprotected and partially aligned with the spacecraft's motion. These special orientations produce long and asymmetric tails in the flux distribution. MEM users tasked with assessing the risk on a surface with a similarly varying exposure pattern should consider requiring a minimum of 1000 state vectors.

\subsection{Future work}

MEM~3 includes an optional standard deviation calculation; the code uses the \cite{welford62} and \cite{west79} algorithms to compute these standard deviations in a single pass. Based on the results presented in this section, we now plan to add a skewness calculation. We will then use the standard deviation and skewness to warn users when their trajectory is likely to produce results coarser than a given precision.

\section{Data and visualizations}
\label{sec:data}

This section describes the full set of data files and graphical visualizations that we have made available in our online meteoroid environment library.

\subsection{Data provided}

In Section~\ref{sec:sample}, we described the steps used to generate or download our example trajectories. However, for those users who have no desire to repeat this process, we provide data files containing the corresponding state vectors. Users can ``click here to download the trajectory.'' These files have been formatted for use with MEM.

We also provide an ``options file'' that documents the set of choices we selected for each MEM run. These options files are in plain text format and can be used as a reference in conjunction with use of the MEM graphic user interface, or they can simply be copied into the user's MEM directory for use with the MEM command-line executable. Either method will result in the user exactly replicating our MEM run results. For the sake of conserving storage space, we have not posted the MEM output files themselves on the MEM library website.

\subsection{Visualizations}
\label{sec:viz}

For each sample environment, we have generated a similar set of visualizations that include [1] three views of the trajectory in an inertial reference frame; [2] three views of the trajectory in a Sun-centered ecliptic reference frame; [3] a bar chart of the flux on flat surfaces with nine specific orientations; [4] a histogram of the impact speed distribution; [5] a histogram of the meteoroid bulk density distribution; and [6] heat maps of meteoroid directionality relative to the spacecraft.

\begin{figure*}
    \includegraphics[scale=0.65]{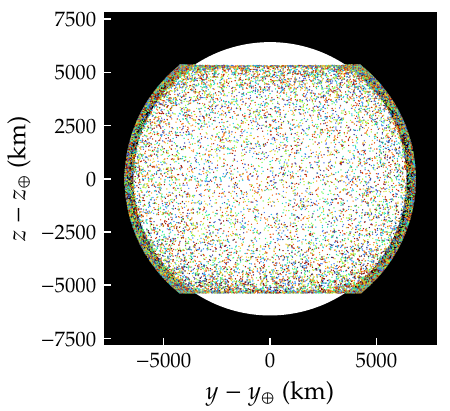}
    \includegraphics[scale=0.65]{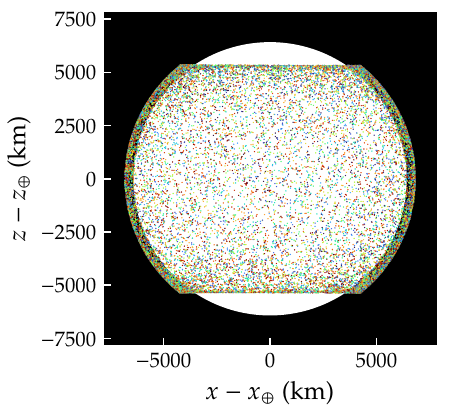}
    \includegraphics[scale=0.65]{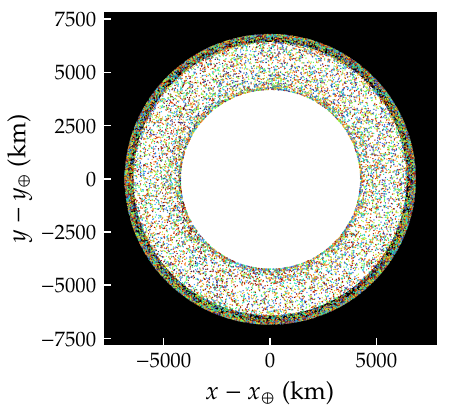}
    \includegraphics[scale=0.65]{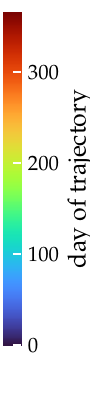}
    \vspace{0.1in}

    \includegraphics[scale=0.65]{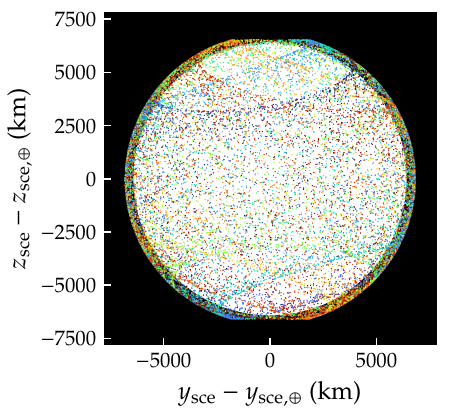}
    \includegraphics[scale=0.65]{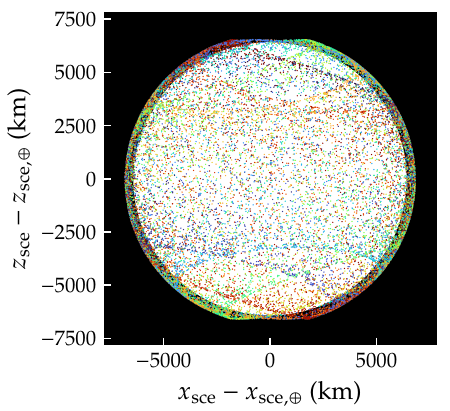}
    \includegraphics[scale=0.65]{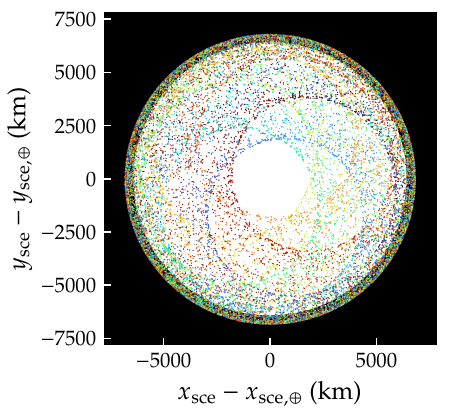}
    \includegraphics[scale=0.65]{iss_cb1.pdf}
    \vspace{0.2in}
    
    \includegraphics[scale=0.65]{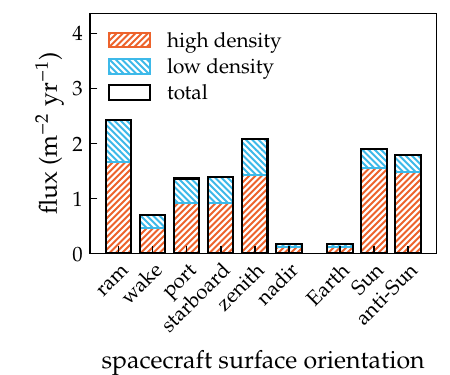}
    \includegraphics[scale=0.65]{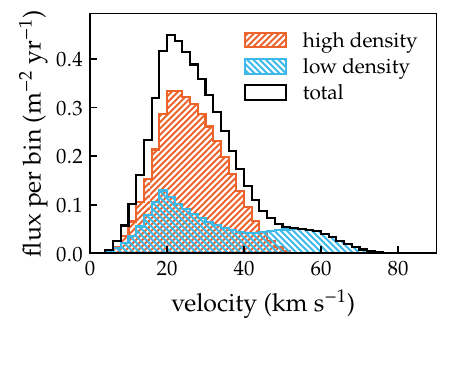}
    \includegraphics[scale=0.65]{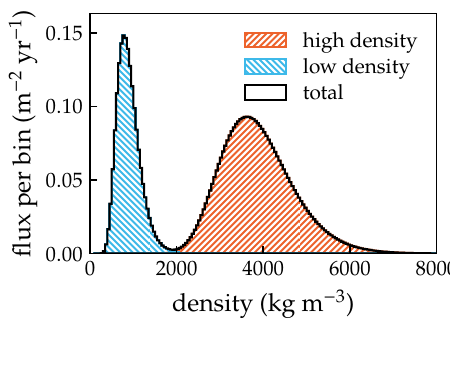}
    \vspace{0.2in}

    \includegraphics[scale=0.64]{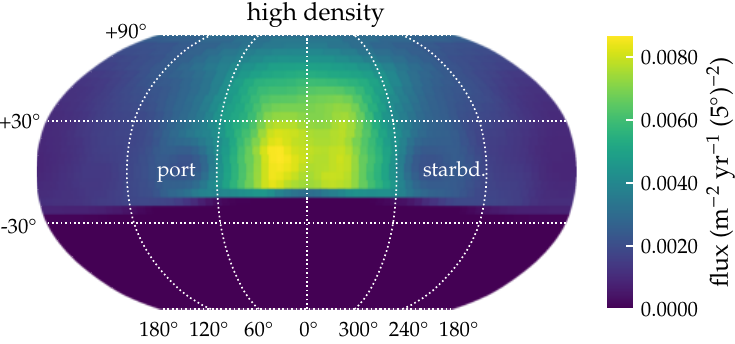}\hfill
    \includegraphics[scale=0.64]{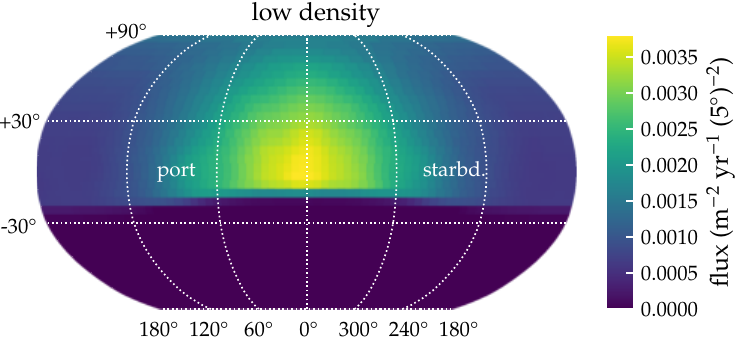}
    \vspace{0.1in}

    \caption{A full suite of visualizations generated for our ISS trajectory. Each plot is discussed in turn in Section~\ref{sec:viz} but, in brief: the top row depicts the trajectory in an inertial frame; the second row depicts the trajectory in a Sun-centered ecliptic frame; the third row shows the microgram-or-larger meteoroid flux as a function of surface orientation, impact speed, and bulk density; and the last row shows the flux as a function of impact angle.}
    \label{fig:issfull}
\end{figure*}

The sample visualizations we present in this paper contain the same information (or a subset of the same information) as those in our online library, but the font size has been reduced. Our online versions are optimized for the web and have large, serif labels and appear in Scalable Vector Graphics (SVG) format. A full set of visualizations for the ISS trajectory is presented in Fig.~\ref{fig:issfull}; we discuss each graphic in this section.

\subsubsection{Trajectory in inertial frame}

In the first row of Fig.~\ref{fig:issfull}, we present a visualization of the ISS trajectory from three angles in an inertial reference frame. In most cases, the axes of the inertial reference frame are aligned with the International Celestial Reference Frame (ICRF), in which the $x$ axis is close to aligned with Earth's dynamical vernal equinox and the $z$ axis is close to aligned with Earth's pole at the J2000 epoch. The sole exception is that of the MAVEN Earth-to-Mars transfer trajectory, which is depicted in an ecliptic reference frame.

In each case, the ``inertial'' frame is centered on the most relevant massive body; in the case of the ISS, this is Earth. Thus, the ISS trajectory is depicted in the 
Earth-centered inertial (ECI) frame. We note that these reference frames are not truly inertial: each planet orbits the Sun and the Sun itself orbits the galactic center. However, we follow the convention of referring to these coordinates as inertial.

\subsubsection{Trajectory in SCE frame}

While ECI coordinates are useful for describing the position of Earth-orbiting spacecraft, they have little relation to the directionality of the meteoroid environment. Sporadic meteoroid radiants are clustered into so-called ``sources'' that maintain the same directionality only when viewed in a Sun-centered ecliptic frame \citep{hawkins56,weiss60,stohl68,jones93,cb08}. This coordinate system is a non-inertial one in which the $+x$ direction points toward the Sun and the $+z$ direction points toward ecliptic north. 

Inertial (ICRF) spacecraft coordinates can be converted to Sun-centered ecliptic coordinates as follows. First, we use the jplephem Python package \citep{jplephem} in conjunction with the DE430 ephemeris \citep{folkner14} to compute the ecliptic coordinates of the central body (in the case of the ISS, Earth).
The solar longitude is then computed as follows:
\begin{align}
    \lambda_\odot &= \text{arctan2}(-y_\text{cb}, \, -x_\text{cb})
\end{align}
We also perform a 23.4$^\circ$ rotation about the $x$-axis to align our spacecraft trajectory with the ecliptic. We then add the spacecraft position to that of the central body to obtain the heliocentric ecliptic position of the spacecraft. We next rotate the heliocentric ecliptic coordinates of both the central body and the spacecraft about the $z$-axis by 
both the central body's heliocentric ecliptic coordinates and the spacecraft's heliocentric coordinates by $\lambda_\odot$. 

Finally, we subtract the coordinates of the central body from that of the spacecraft. The resulting coordinates give the position of the spacecraft relative to the central body in a coordinate frame that is aligned with the ecliptic and has an $x$-axis that always points toward the Sun. The ``Sun-centered ecliptic'' name is somewhat confusing in this case, as the coordinate frame is \emph{positionally} centered on the central body, but the orientation of the $x$-axis is centered on the Sun. In our graphics, we use the subscript ``SCE'' to indicate that the $x$-axis is oriented towards the Sun, and we subtract $x_\oplus$ or $x_{\text{SCE},\oplus}$, for instance, to indicate that position is measured relative to Earth. 

The second row of Fig.~\ref{fig:issfull} displays the position of the ISS relative to Earth in an SCE coordinate frame. We see that we have sampled a wide range of possible Sun-Earth-spacecraft positions and thus will have a good average meteoroid environment description for spacecraft on this type of orbit.

\subsubsection{Surface flux bar chart}
\label{sec:bar}

We next present a bar chart that provides the flux of microgram-or-larger meteoroids onto different surfaces of a cubic spacecraft (third row of Fig.~\ref{fig:issfull}). We assume that these surfaces maintain the same orientation relative to the spacecraft's orbital motion; for instance, we assume that the ``ram'' surface always faces in the direction of the spacecraft's orbital velocity. 

Readers may wish to consult Section~4.1.1 of \cite{MEM3TM} for a more in-depth discussion of this coordinate frame, which is referred to in that document as ``body-fixed.'' We would like to acknowledge here that ``body-fixed'' is not the standard term for this reference system; when the central body is a planet or Moon, MEM's body-fixed coordinate frame is better known as a VNB (velocity/normal/binormal) reference frame.\footnote{https://gmat.sourceforge.net/docs/R2016a/help.html} 

The first plot in the third row of Fig.~\ref{fig:issfull} provides the flux of microgram-or-larger meteoroids on surfaces facing along each positive and negative axis of the body-fixed or VNB coordinate system. We also include the flux on surfaces facing Earth and towards or away from the Sun; these fluxes may be relevant for assessing the risk posed to communications equipment or solar panels. 

MEM includes three dynamical populations with two different bulk density distributions. The high-density population originates from Jupiter-family comets and is also called the helion/anti-helion population, and the low-density population consists of the apex and toroidal populations which originate from long-period and Halley-type comets, respectively \citep{jones93,2017MNRAS.472.3833M,2020JSpRo..57..160M}. In these visualizations, we stack the flux from the two distributions so that users can view the total flux.

In Section~\ref{sec:select}, we made multiple comments about how certain spacecraft surfaces can be relatively protected from the meteoroid environment due to planetary shielding. The surface flux bar chart can be used to visually assess how strong this effect is; for instance, Fig.~\ref{fig:issfull} shows us that the flux on the nadir-facing surface of ISS is roughly an order of magnitude lower than that on the zenith-facing surface.

\subsubsection{Mass distribution}
\label{sec:mass}

Readers may notice that we do not include a mass distribution plot. The reason is that the mass distribution does not vary between trajectories or meteoroid populations, and therefore always resembles Fig.~1 of \cite{MEM3TM} in shape. All runs in our library use the default minimum mass of 1~$\mu$g \citep[see Sec.~3.2.3 of][]{MEM3TM}.

\subsubsection{Speed distribution}

The third row of Fig.~\ref{fig:issfull} also includes a histogram of meteoroid speeds (center plot). These speeds are relative to the spacecraft; the spacecraft's velocity has been taken into account. We present the speed distribution for each density population as well as the overall speed distribution. In this case, the speed distribution is that over the entire spacecraft, assuming that it does not maintain any particular orientation. Orientation-specific speed distributions are available as a MEM output.

\subsubsection{Density distribution}

The last plot in the third row of Fig.~\ref{fig:issfull} displays the distribution of bulk density for each meteoroid population. Bulk density plays a role in certain ballistic limit equations \citep[such as the modified Cour-Palais BLE;][]{hayashida91}.

\subsubsection{Directionality maps}

The last row of Fig.~\ref{fig:issfull} contains two directional flux maps. These maps use a color scale to show how the flux (per 25-square-degree bin) varies with impact angle. The azimuthal angle measures the angle counterclockwise from the $+x$-axis within the $x$-$y$ plane, and the elevation angle measures the angular offset from the $x$-$y$ plane. Notice that we have centered the azimuthal axis on the direction of motion and reversed the azimuthal axis: this results in an ``inside-out'' view of meteoroid directionality in which port appears to the left of the direction of motion.

In this case, we see that the ISS encounters little-to-no flux from ``below.'' This is because the ISS has a fairly low altitude: 400~km above Earth's surface and 300~km above the altitude at which Earth's atmosphere is capable of blocking meteoroids. Thus, the ISS is protected from a large portion of the meteoroid environment due to Earth's shielding effects.

\section{Summary}
\label{sec:summary}

This paper announces the availability of a new online meteoroid environment library.\footnote{https://fireballs.ndc.nasa.gov/mem/library/} This library includes a meteoroid environment description for at least one spacecraft orbiting every major body in the inner solar system as well as one planetary transfer trajectory and two Lagrange halo orbits. We provide visualizations of the meteoroid flux generated with the Meteoroid Engineering Model, version 3 (MEM~3), and our library includes the input files needed to replicate these runs.

Our hope is that this online library will be a useful tool to aerospace engineers who are conducting meteoroid risk assessments. Our graphics provide a visual guide to the meteoroid environment encountered by spacecraft; users might also choose to compare flux plots to get an idea of how much meteoroid flux, directionality, and speed varies between different types of trajectories or near different planets. The included input files can be used as test runs for a new installation of MEM, and the resulting output files may be useful to users developing risk analysis tools. The output may even be used for \emph{preliminary} risk assessments, such as obtaining a rough estimate of the meteoroid flux encountered by an Earth-orbiting satellite on a Sun-synchronous orbit. Please note, however, that these example runs should not be substituted for a mission-specific trajectory and MEM run in a final or ``official'' risk assessment.

It is possible that our library may also have scientific applications. For instance, the James Webb Space Telescope (JWST) has experienced a number of meteoroid strikes since its deployment in early 2022.\footnote{https://blogs.nasa.gov/webb/2022/06/08/webb-engineered-to-endure-micrometeoroid-impacts/} The JWST entry in our library allows meteoroid researchers to examine the pattern of meteoroid encounters predicted by MEM for JWST and to compare it with that of other models. 

In the course of generating these trajectories and MEM runs, we have also developed a method to determine the trajectory size, in number of state vectors, needed to achieve a desired flux resolution. Users can adopt the approach outlined in Section~\ref{sec:method} to ensure that their trajectories are just detailed enough for their needs. This helps users avoid runs that have too low a resolution and also avoid the excessively time-consuming runs that result from trajectories with far too many state vectors. It should be noted, however, that this precision does \emph{not} reflect the uncertainty in the meteoroid environment, which is believed to be roughly a factor of 2-3 near 1~au \citep[see][for additional discussion of environment uncertainties]{2020JSpRo..57..160M}. A recommended implementation of the estimated environment uncertainty will be the subject of a future paper.

\section*{Acknowledgments}

This work was supported in part by NASA contract
80MSFC18C0011 and the NASA Marshall Space Flight Center's internship program.

\bibliography{local}

\end{document}